\documentclass[a4paper]{article}

\usepackage{INTERSPEECH2018}
\usepackage{xcolor}

\title{Differentiable Supervector Extraction for Encoding Speaker and Phrase Information in Text Dependent Speaker Verification}

\name{Victoria Mingote, Antonio Miguel, Alfonso Ortega, Eduardo Lleida}
\address{
  ViVoLab, Arag\'{o}n Institute for Engineering Research (I3A), University of Zaragoza, Spain
  }
\email{\{vmingote,amiguel,ortega,lleida\}@unizar.es}
\begin{document}
\maketitle
\begin{abstract}
In this paper, we propose a new differentiable neural network alignment mechanism for text-dependent speaker verification which uses alignment models to produce a supervector representation of an utterance.
Unlike previous works with similar approaches, we do not extract the embedding of an utterance from the mean reduction of the temporal dimension.
Our system replaces the mean by a phrase alignment model to keep the temporal structure of each phrase which is relevant in this application since the phonetic information is part of the identity in the verification task.
Moreover, we can apply a convolutional neural network as front-end, and thanks to the alignment process being differentiable, we can train the whole network to produce a supervector for each utterance which will be discriminative with respect to the speaker and the phrase simultaneously.
As we show, this choice has the advantage that the supervector encodes the phrase and speaker information providing good performance in text-dependent speaker verification tasks.
In this work, the process of verification is performed using a basic similarity metric, due to simplicity, compared to other more elaborate models that are commonly used.
The new model using alignment to produce supervectors was tested on the RSR2015-Part I database for text-dependent speaker verification, providing  competitive results compared to similar size networks using the mean to extract embeddings.
\end{abstract}
\footnote{\color{red}{Cite as: Mingote, V., Miguel, A., Ortega, A., Lleida, E. (2018) Differentiable Supervector Extraction for Encoding Speaker and Phrase Information in Text Dependent Speaker Verification. Proc. IberSPEECH 2018, 1-5, DOI: 10.21437/IberSPEECH.2018-1.}}
\vspace{-0.1cm}
\section{Introduction}
Recently, techniques based on discriminative deep neural networks (DNN) have achieved a substantial success in many speaker verification tasks. 
These techniques follow the philosophy of the state-of-the-art face verification systems \cite{Taigman2014DeepFace:Verification}\cite{7298682} where embeddings are usually extracted by reduction mechanisms and the decision process is based on a similarity metric \cite{nguyen2010cosine}.
Unfortunately, in text-dependent tasks this approach does not work efficiently since the pronounced phrase is part of the identity information \cite{Liu2015DeepVerification}\cite{Malykh2017OnTask}.
A possible cause of the imprecision in text-dependent tasks could be derived from using the mean as a representation of the utterance as we show in the experimental section. 
To solve this problem, this paper shows a new architecture which combines a deep neural network with a phrase alignment method used as a new internal layer to maintain the temporal structure of the utterance.
As we will show, it is a more natural solution for the text-dependent speaker verification, since the speaker and phrase information can be encoded in the supervector thanks to the neural network and the specific states of the supervector.

In the context of text-independent speaker verification tasks, the baseline system based on i-vector extraction and Probabilistic Linear Discriminant Analysis (PLDA) \cite{kenny2008study}\cite{dehak2011front} are still among the best results of the state-of-the-art. 
The i-vector extractor represents each utterance in a low-dimensional subspace called the total variability subspace as a fixed-length feature vector and the PLDA model produces the verification scores.
However, as we previously mentioned, many improvements on this baseline system have been achieved in recent years by progressively substituting components of the systems by DNNs, thanks to their larger expressiveness and the availability of bigger databases.
Examples of this are the use of DNN bottleneck representations as features replacing or combined with spectral parametrization \cite{lozano2016analysis}, training DNN acoustic models to use their outputs as posteriors for alignment instead of GMMs in i-vector extractors \cite{lei2014novel}, or replacing PLDA by a DNN \cite{ghahabi2014deep}.
Other proposals similar to face verification architectures have been more ambitious and have trained a discriminative DNN for multiclass classifying and then extract embeddings by reduction mechanisms \cite{bhattacharya2017deep} \cite{snyder2017deep}, for example taking the mean of an intermediate layer named usually bottleneck layer.
After that embedding extraction, the verification score is obtained by a similarity metric such as cosine similarity \cite{bhattacharya2017deep}.

The application of DNNs and the same techniques as in text-independent models for text-dependent speaker verification tasks  has produced mixed results. 
On the one hand, specific modifications of the traditional techniques have been shown successful for text-dependent tasks such as i-vector+PLDA \cite{zeinali2017hmm}, DNNs bottleneck as features for i-vector extractors \cite{zeinali2016deep} or posterior probabilities for i-vector extractors \cite{zeinali2016deep}\cite{dey2016deep}.
On the other hand, speaker embeddings obtained directly from a DNN have provided good results in tasks with large amounts of data and a single phrase \cite{Heigold2016End-to-endVerification} but they have not been as effective in tasks with more than one pass phrase and smaller database sizes \cite{Liu2015DeepVerification}\cite{Malykh2017OnTask}.
The lack of data in this last scenario may lead to problems with deep architectures due to overfitting of models.

Another reason that we explore in the paper for the lack of effectiveness of these techniques in general text-dependent tasks is that the phonetic content of the uttered phrase is relevant for the identification.
State-of-art text-independent approaches to obtain speaker embeddings from an utterance usually reduce temporal information by pooling and by calculating the mean across frames of the internal representations of the network.
This approach may neglect the order of the phonetic information because in the same phrase the beginning of the sentence may be totally different from what is said at the end.
An example of this is the case when the system asks the speaker to utter digits in some random order.
In that case a mean vector would fail to capture the combination of phrase and speaker.
Therefore one of the objectives of the paper is to show that it is important to keep this phrase information for the identification process, not just the information of who is speaking.

In previous works we have developed systems that need to store a model per user which were adapted from a universal background model and the evaluation of the trial was based on a likelihood ratio \cite{Miguel2014FactorRecognition}\cite{Miguel2017TiedRecognition}.
One of the drawbacks of this approach is the need to store a large amount of data per user and the speed of evaluation of trials, since likelihood expressions were dependent on the frame length.
In this paper, we focus on systems using a vector representation of a trial or a speaker model.
We propose a new approach that includes alignment as a key component of the mechanism to obtain the vector representation from a deep neural network.
Unlike previous works, we substitute the mean of the internal representations across time which is used in other neural network architectures \cite{Liu2015DeepVerification}\cite{Malykh2017OnTask} by a frame to state alignment to keep the temporal structure of each utterance.
We show how the alignment can be applied in combination with a DNN acting as a front-end to create a supervector for each utterance. 
%
%
As we will show, the application of both sources of information in the process of defining the supervector provides better results in the experiments performed on RSR2015 compared to previous approaches.

This paper is organized as follows. In Section 2 we present our system and especially the alignment strategy developed. Section 3 presents the experimental data. Section 4 explains the results achieved. Conclusions are presented in Section5.
\vspace{-0.1cm}
\section{Deep neural network based on alignment}
In view of the aforementioned imprecisions in the results achieved in previous works for this task with only DNNs and a basic similarity metric, we decided to apply an alignment mechanism due to the importance of the phrases and their temporal structure in this kind of tasks.
Since same person does not always pronounce one phrase at the same speed or in the same way due to differences in the phonetic information, it is usual that there exists an articulation and pronunciation mismatch between two compared speech utterances even from the same person.

\begin{figure}[th]
  \centering
  \includegraphics[width=0.77\linewidth]{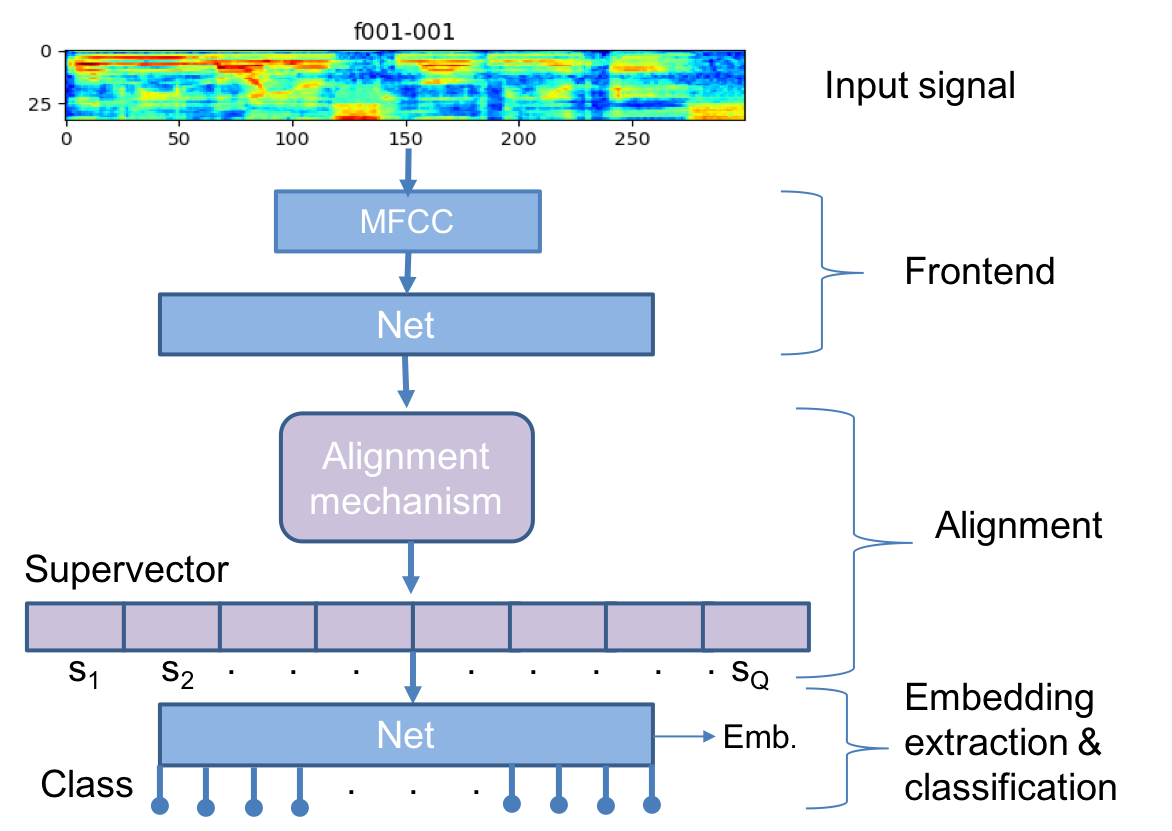}
  \caption{Differentiable neural network alignment mechanism based on alignment models. The supervector is composed of Q vectors $s_{Q}$  for each state.} 
  \label{fig1}
  \vspace{-0.4cm}
\end{figure}

In Fig.~\ref{fig1} we show the overall architecture of our system, where the mean reduction to obtain the vector embedding before the backend is substituted by the alignment process to finally create a supervector by audio file. 
This supervector can be seen as a mapping between an utterance and the state components of the alignment, which allows to encode the phrase information.
For the verification process, once our system is trained, one supervector is extracted for each enroll and test file, and then a cosine metric is applied over them to achieve the verification scores.
\vspace{-0.2cm}

\subsection{Alignment mechanism}
In this work, we select a Hidden Markov Model (HMM) as the alignment technique in all the experiments, but other posibilities could be to select Gaussian Mixture Model (GMM) or DNN posteriors. 
In text-dependent tasks we know the phrase transcription which allows us to construct a specific left-to-right HMM model for each phrase of the data and obtain a Viterbi alignment per utterance.

One reason to employ a phrase HMM alignment was due to its simplicity for training independent HMM models for different phrases used to develop our experiments without the need of phonetic information for training. 
Another reason was that using the decoded sequence provided by the Viterbi algorithm in a left-to-right architecture it is ensured that each state of the HMM corresponds to at least one frame of the utterance, so no state is empty.

The process followed to add this alignment to our system is detailed below.
Once models for alignment are trained, a sequence of decoded states \textbf{$\gamma$}=$(q_{1},...,q_{t})$ where \textit{$q_{t}$} indicates the decoded state at time t with $\textit{$q_{t}$}\in\{1, ..., Q\}$ is obtained. 
Before adding these vectors to the neural network they are preprocessed and converted into a matrix with ones and zeros in function of its correspondences with the states which makes possible to use them directly inside of the neural network. 
In this way, we put ones at each state according to the frames that belong to this state as a result of this process, we have the alignment matrix $\textit{A}\in\mathbb{R}^{T \times Q}$ with its components $\textit{$a_{tq_{t}}$=1}$ and $\textit{$\sum_{q}a_{tq}$=1}$ which means that only one state is active at the same time.

For example, if we train an HMM model with 4 states and we obtain a vector $\gamma$ and apply the previous transformation, the resultant matrix $A$ would be:
\begin{equation}
\gamma=[1,1,1,2,2,3,3,4]\rightarrow
A=
\left(
\begin{array}{cccc}
1 & 0 & 0& 0 \\
1 & 0 & 0 & 0 \\
1 & 0 & 0 & 0 \\
0 & 1 & 0 & 0 \\
0 & 1 & 0& 0 \\
0 & 0 & 1 & 0 \\
0 & 0 & 1 & 0 \\
0 & 0 & 0 & 1
\end{array}
\right) 
\end{equation}

\begin{figure}[t]
	\centering
	\includegraphics[width=0.9\linewidth]{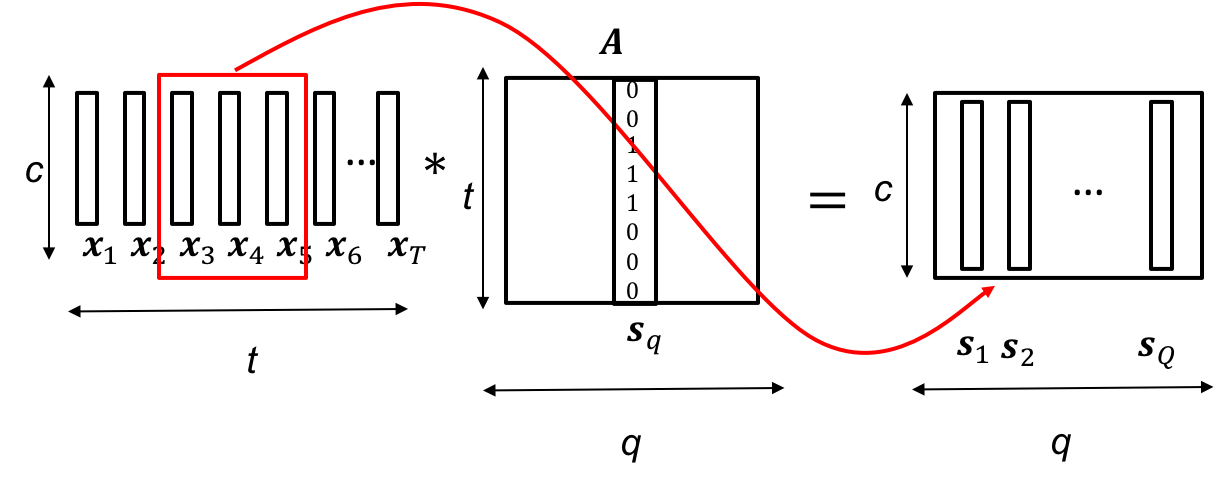}
	\caption{Process of alignment, the input signal x is multiplied by an alignment matrix A to produce a matrix with vectors $s_{Q}$ which are then concatenated to obtain the supervector.} 
    \label{fig2}
    \vspace{-0.4cm}
\end{figure}

After this process, as we show in Fig.~\ref{fig2}, we added this matrix to the network as a matrix multiplication like one layer more, thanks to the expression as a matrix product it is easy to differentiate and this enables to backpropagate gradients to train neural network as usual.
This matrix multiplication allows assigning the corresponding frames to each state resulting in a supervector. Then, the speaker verification is performed with this supervector.
The alignment as a matrix multiplication can be expressed as a function of the input signal to this layer \textbf{$x_{ct}$} with dimensions \textit{(c $\times$ t)} and matrix of alignment of each utterance \textbf{$A$} with dimensions \textit{(t $\times$ q)}: 
\begin{equation}
s_{cq}=\frac{\sum_{t}x_{ct}\cdot a_{tq}}{\sum_{t}a_{tq}}
\end{equation}
where $s_{cq}$ is the supervectors with dimensions \textit{(c $\times$ q)}, where there are \textit{q} state vectors of dimension \textit{c} and we normalize with the number of times state \textit{q} is activated.

\subsection{Deep neural network architecture}
As a first approximation to check that the previous alignment layer works better than extracting the embedding from the mean reduction, we apply this mentioned layer directly on the input signal over the acoustic features thus we obtain the traditional supervector.
However, we expect to improve this baseline result, so we propose to add some layers as front-end previous to the alignment layer and train them in combination with the alignment mechanism.

For deep speaker verification some simple architectures with only dense layers \cite{Liu2015DeepVerification} have been proposed. 
However, lately it has been tried to employ deep neural networks as Residual CNN Networks \cite{Malykh2017OnTask} but in text-dependent task it has not achieved the same good results as previous simple approaches. 

In our network we propose a straightforward architecture with only a few layers which include the use of 1-dimension convolution (1D convolution) layers instead of dense layers or 2D convolution layers as in other works.
Our proposal is to operate in the temporal dimension to add context information to the process and at the same time the channels are combined at each layer. The context information which is added depends on the size of the kernel used in convolution layer. 

To use this type of layer, it is convenient that the input signals have the same size to concatenate them and pass to the network. For this reason, we apply a transformation to interpolate or fill with zeros the input signals to have all of them with the same dimensions.

\begin{figure}[th]
	\centering
	\subfigure[Operation with 1D Convolution]{
 	\includegraphics[width=0.8\linewidth]
{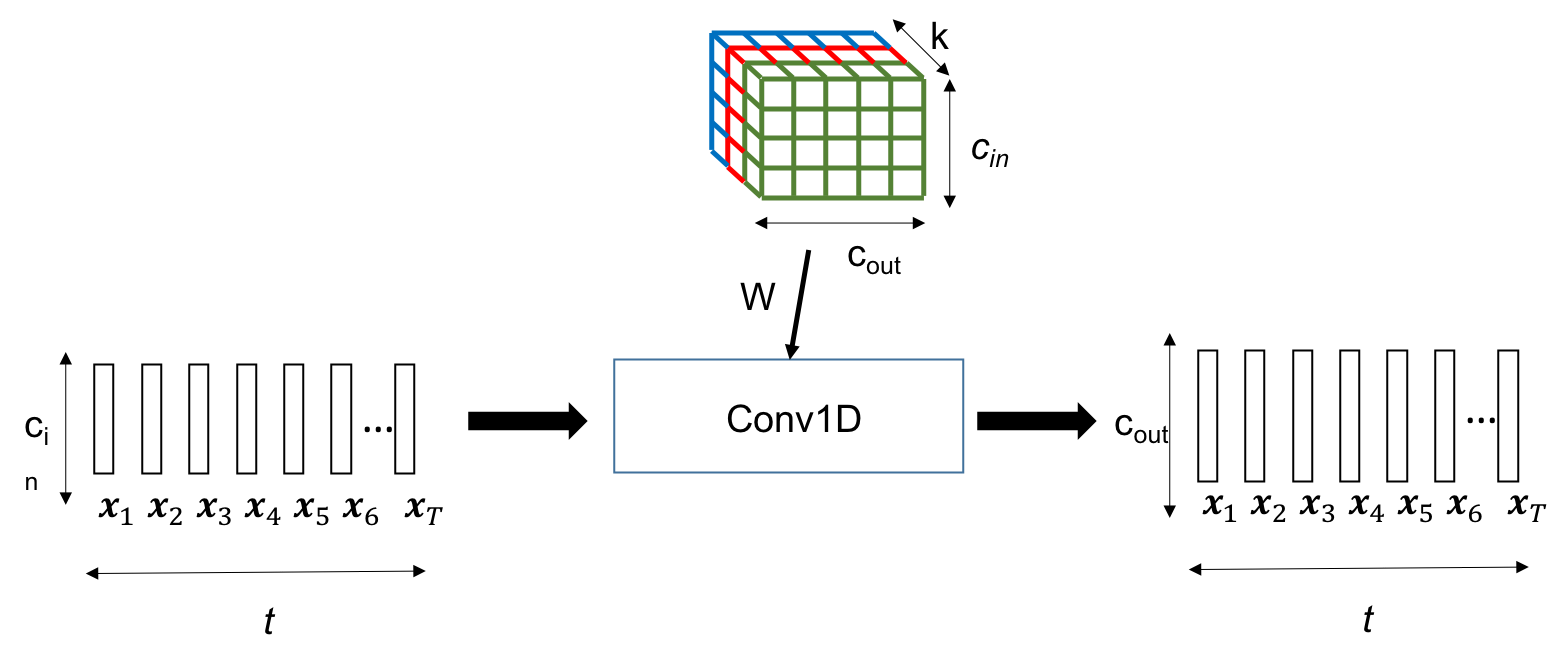}
	\label{fig3:a}}
    \centering
	\subfigure[Example of the convolution operation]{
	\includegraphics[width=0.8\linewidth]{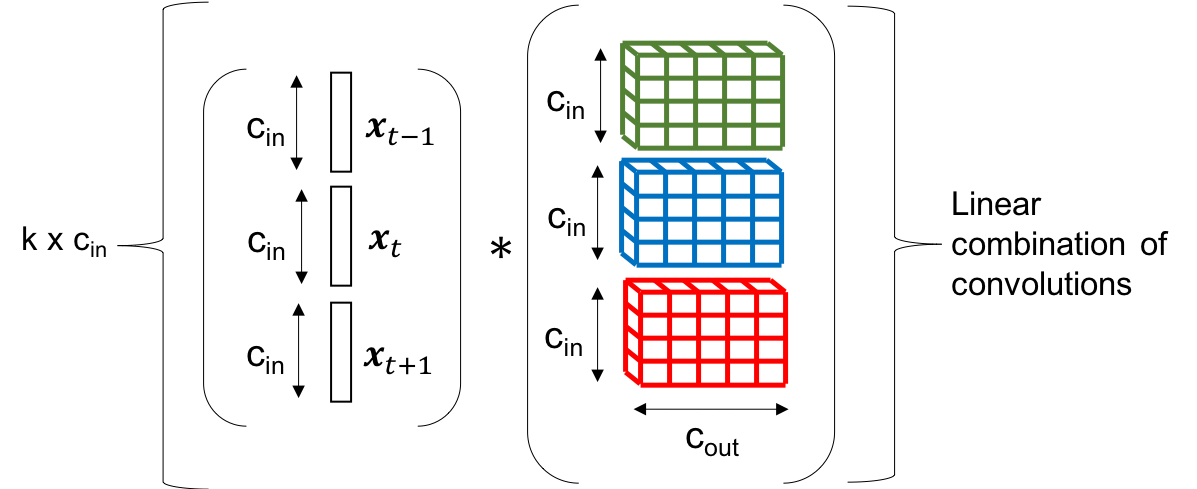}
    \label{fig3:b}}
    \caption{Operation with 1D Convolution layers, \ref{fig3:a} general pipeline of this operation. \ref{fig3:b} example of how k context frames from input are multiplied by the weight matrix W and the output is equivalent to a linear combination of convolutions.} 
    \label{fig3}
    \vspace{-0.4cm}
\end{figure}

The operation of the 1D convolution layers is depicted in Fig.~\ref{fig3}, the signal used as layer input and its context, the previous frames and the subsequent frames, are multiplied frame by frame with the corresponding weights.
The result of this operation for each frame is linearly combined to create the output signal.
\vspace{-0.2cm}
\section{Experimental Data}
In all the experiments in this paper, we used the RSR2015 text-dependent speaker verification dataset \cite{Larcher2014Text-dependentRSR2015}. This dataset consists of recordings from 157 male and 143 female. There are 9 sessions for each speaker pronouncing 30 different phrases. Furthermore, this data is divided into three speaker subset: background (bkg), development (dev) and evaluation (eval). We develop our experiments in Part I of this data set and we employ the bkg and dev data (194 speakers, 94 female/100 male) for training. The evaluation part is used for enrollment and trial evaluation. 
\vspace{-0.2cm}

\section{Results}
In our experiments, we do not need the phrase transcription to obtain the corresponding alignment, because one phrase dependent HMM model has been trained with the background partition using a left-to-right model of 40 states for each phrase.
With these models we can extract statistics from each utterance of the database and use this alignment information inside our DNN architecture.
As input to the DNN, we employ 20 dimensional Mel-Frequency Cepstral Coefficients (MFCC) with their first and second derivatives as features for obtaining a final input dimension of 60.
On these input features we apply a data augmentation method called Random Erasing \cite{zhong2017random}, which helps us to avoid overfitting in our models due to lack of data in this database. 

On the other hand, the DNN architecture consists of the front-end part in which several different configurations of layers have been tested as we will detail in the experiments, and the second part of the architecture which is an alignment based on HMM models.
Finally, we have extracted supervectors as a combination of front-end and alignment with a flatten layer and with them we have obtained speaker verification scores by using a cosine similarity metric without any normalization technique.

A set of experiments was performed using Pytorch \cite{paszke2017automatic} to evaluate our system. We compare a front-end with mean reduction with similar philosophy as \cite{Liu2015DeepVerification}\cite{Malykh2017OnTask}  to the feature input directly or a front-end both followed by the HMM alignment. 
In the part of the front-end, we implemented 3 different layer configurations: one convolutional layer with a kernel of dimension 1 equivalent to a dense layer but keeping the temporal structure and without adding context information, one convolutional layer with a kernel of dimension 3, and three convolutional layers with a kernel of dimension 3.

In Table 1 we show equal error rate (EER) results with the different architectures trained on the background subset for female, male and both partitions together. 
We have found that, as we expected, the first approach with mean reduction mechanism for extracting embeddings does not perform well for this text-dependent speaker verification task. It seems that this type of embeddings do not represent correctly the information to achieve discrimination between the correct speaker and phrase both simultaneously.
Furthermore, we show how changing the typical mean reduction for a new alignment layer inside the DNN achieves a relative improvement of 91.62\% in terms of the EER\%.

\begin{figure}[th]
	\centering
\vspace{-0.2cm}	\includegraphics[width=0.75\linewidth]{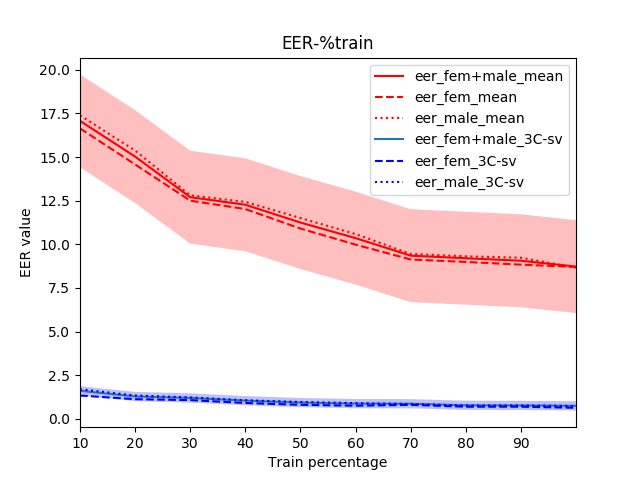}
	\caption{Results of EER\% varying train percentage where standard deviation is shown only for both gender independent results.} 
    \label{fig4}
\end{figure}

\begin{table}[t]
  \caption{Experimental results on RSR2015 part I \cite{Larcher2014Text-dependentRSR2015} eval subset, where EER\% is shown. These results were obtained by training only with bkg subset.}
  \label{tab:example}
  \centering
  \resizebox{0.475\textwidth}{!} {
  \begin{tabular}{ rrrrr}
    \toprule
    \multicolumn{1}{c}{\textbf{Architecture}} &
    \multicolumn{1}{c}{} &
    \multicolumn{1}{c}{\textbf{Fem}} & 
    \multicolumn{1}{c}{\textbf{Male}} & 
    \multicolumn{1}{c} {\textbf{Fem+Male}} \\
    \cline{1-2}
    \multicolumn{1}{c}{Layers}&
    \multicolumn{1}{c}{Kernel}\\
    
    \midrule
    $FE:3C+mean$ &$3$ & $11.20\%$ & $12.13\%$& $11.70\%$~~~ \\
    $Signal+alig.$ &$-$ & $1.43\%$  & $1.37\%$& $1.54\%$~~~ \\
    $FE:1C+alig.$ &$1$  & $1.16\%$  & $0.98\%$& $1.56\%$~~~ \\
    $FE:1C+alig.$ &$3$  & $1.04\%$  & $0.77\%$& $1.20\%$~~~ \\
    $FE:3C+alig.$ &$3$  & $0.86\%$  & $1.00\%$& $0.98\%$~~~ \\
   
    \bottomrule
  \end{tabular}}
  
\end{table}
Nevertheless, these EER results were still quite high, so we decided that the results can be improved training with background and develop subsets together.  
In Table 2, we can see that if we use more data for training our systems, we achieve better performance especially in deep architectures with more than one layer, this improvement is observed for both architectures.
This fact remarks the importance of having a large amount of data to be able to train deep architectures.
In addition, we performed an experiment to illustrate this effect in Fig.\ref{fig4} where we show how if we increase little by little the amount of data used to train, the results progressively improve although we can see that the alignment mechanism makes the system more robust to training data size.  

\begin{table}[th]
  \caption{Experimental results on RSR2015 part I \cite{Larcher2014Text-dependentRSR2015} eval subset, showing EER\%. These results were obtained by training with bkg+dev subsets.}
  \label{tab:example}
  \centering
  \resizebox{0.475\textwidth}{!} {
  \begin{tabular}{ r r r r r}
    \toprule
    \multicolumn{1}{c}{\textbf{Architecture}} &
    \multicolumn{1}{c}{} &
    \multicolumn{1}{c}{\textbf{Fem}} & 
    \multicolumn{1}{c}{\textbf{Male}} & 
    \multicolumn{1}{c} {\textbf{Fem+Male}} \\
    \cline{1-2}
    \multicolumn{1}{c}{Layers}&
    \multicolumn{1}{c}{Kernel}\\
    \midrule
    $FE:3C+mean$&$3$& $9.11\%$ & $8.66\%$& $8.87\%$~~~ \\
    $Signal+alig.$&$-$& $1.43\%$  & $1.37\%$& $1.54\%$~~~ \\
    $FE:1C+alig.$&$1$ & $1.17\%$  & $0.98\%$& $1.55\%$~~~ \\
    $FE:1C+alig.$&$3$ & $1.07\%$  & $0.78\%$& $1.24\%$~~~ \\
    $FE:3C+alig.$&$3$ & $0.58\%$  & $0.70\%$& $0.72\%$~~~ \\
   
    \bottomrule
  \end{tabular}} 
  \vspace{-0.4cm}
\end{table}

\begin{figure}[th]
	\centering
	\subfigure[Mean embeddings]{
 	\includegraphics[width=0.475\linewidth]
{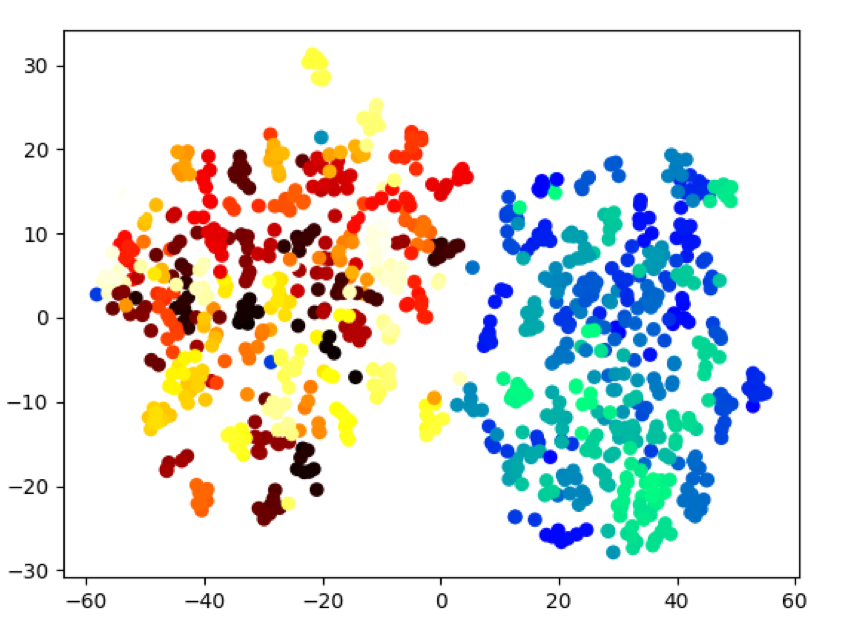}
	\label{fig5:a}}
	\subfigure[Supervectors]{
	\includegraphics[width=0.475\linewidth]	{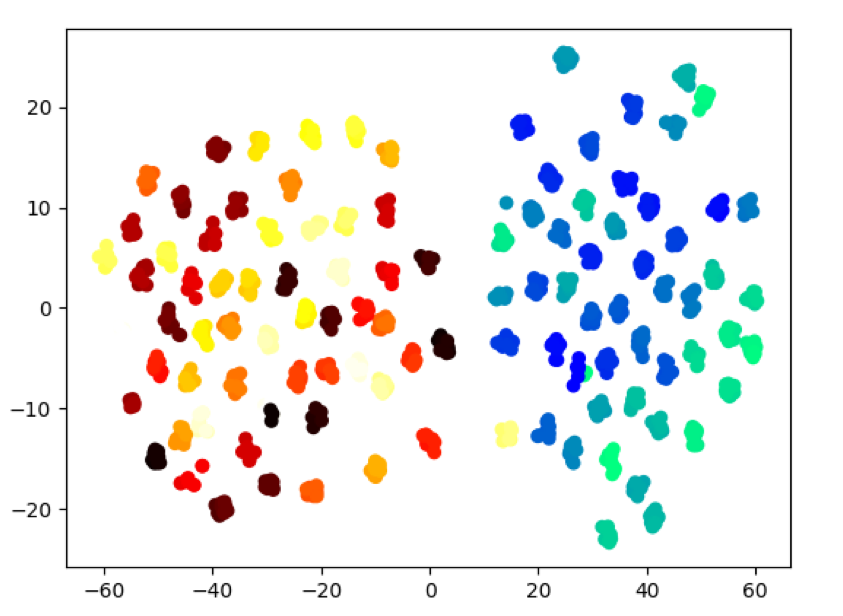}
    \label{fig5:b}}
    \caption{Visualizing Mean embeddings vs Supervectors for 1 phrase from male+female using t-SNE, where female is marked by cold color scale and male is marked by hot color scale.} 
    \label{fig5}
    \vspace{-0.4cm}
\end{figure}

For illustrative purposes, we also represent our high-dimensional supervectors in a two-dimensional space using t-SNE \cite{maaten2008visualizing} which preserves distances in a small dimension space.
In Fig.\ref{fig5:a}, we show this representation for the architecture which uses the mean to extract the embeddings, while in Fig.\ref{fig5:b} we represent the supervectors of our best system. 
As we can see in the second system the representation is able to cluster the examples from the same person, whereas in the first method is not able to cluster together examples from the same person.
On the other hand, in both representations data are auto-organized to show on one side examples from female identities and on the other side examples from male identities.

\begin{figure}[th]
	\centering
	\subfigure[Mean embeddings]{
 	\includegraphics[width=0.475\linewidth]
{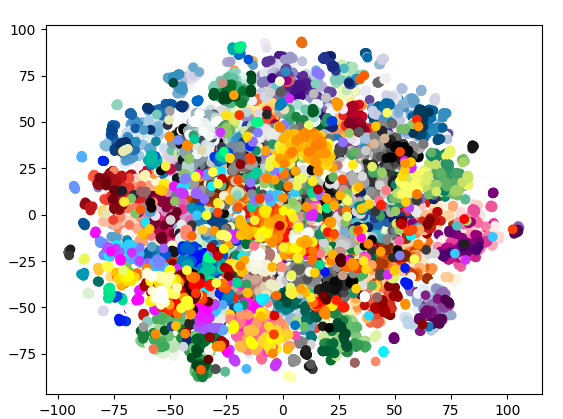}
	\label{fig6:a}}
	\subfigure[Supervectors]{
	\includegraphics[width=0.475\linewidth]	{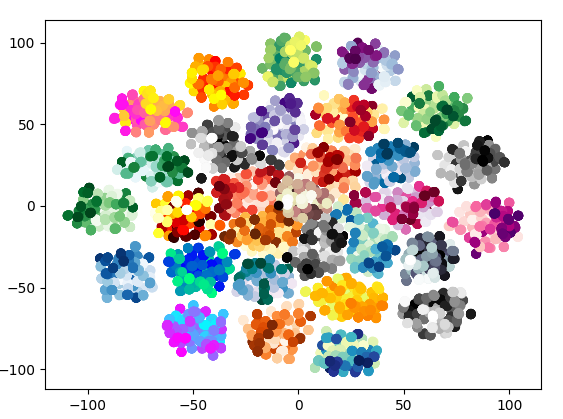}
    \label{fig6:b}}
    \caption{Visualizing Mean embeddings vs Supervectors for 30 phrases from female using t-SNE. Each phrase is marked by one different color scale.} 
    \label{fig6}
\end{figure}

Furthermore, we illustrate in Fig.\ref{fig6} the same representation in the previous figure, however in this case we represent the embeddings and the supervectors of the thirty phrases from female identities. 
With this depiction we checked something that we had already observed in the previous verification experiments since the embeddings from mean architecture are not able to separate between same identity with different phrase and same identity with the same phrase which is the base of text-dependent speaker verification task.
\vspace{-0.5cm}
\section{Conclusions}
In this paper we present a new method to add a new layer as an alignment inside of the DNN architectures for encoding meaningful information from each utterance in a supervector, which allows us to conserve the relevant information that we use to verify the speaker identity and the correspondence with the correct phrase. 
We have evaluated the models in the text-dependent speaker verification database RSR2015 part I. Results confirm that the alignment as a layer within the architecture of DNN is an interesting line since we have obtained competitive results with a straightforward and simple alignment technique which has a low computational cost, so we can achieve better results with other more powerful techniques.

\section{Acknowledgements}
This work has been supported by the Spanish Ministry of Economy and Competitiveness and the European Social Fund through the project TIN2017-85854-C4-1-R, by Gobierno de Arag\'{o}n/FEDER (research group  T36\_17R) and by Nuance Communications, Inc. We gratefully acknowledge the support of NVIDIA Corporation with the donation of the Titan Xp GPU used for this research.

\bibliographystyle{IEEEtran}

\bibliography{mybib}


\end{document}